\documentclass[11pt]{article}
\usepackage[a4paper,margin=1in]{geometry}
\usepackage{amsmath,amsthm,amssymb,mathtools,bm}
\usepackage{microtype}
\usepackage{natbib}
\usepackage{hyperref}
\hypersetup{colorlinks=true,linkcolor=blue,citecolor=blue,urlcolor=blue}
\usepackage{tikz}
\usetikzlibrary{
  arrows.meta,        % for -{Latex}
  positioning,        % for relative positioning like below=2pt of ...
  shapes.geometric,   % for box styles like rounded corners
  calc,
  decorations.pathreplacing % for braces
}
\usepackage{subcaption}

\usepackage[normalem]{ulem} % for \sout
\usepackage{xcolor}
\usepackage{soul}

\newcommand{\responseplus}[1]{{#1}}

\usepackage[utf8]{inputenc}
\usepackage[T1]{fontenc}

\usepackage{amsmath,amssymb,amsthm,amsfonts}
\usepackage{graphicx}
\usepackage{booktabs}
\usepackage{hyperref}
\hypersetup{colorlinks=true,linkcolor=blue,citecolor=blue,urlcolor=blue}

\newtheorem{lemma}{Lemma}[section]
\newtheorem{remark}{Remark}[section]

\newtheorem{proposition}{Proposition}[section]
\newtheorem{corollary}{Corollary}[section]

\title{Perceived Fairness in Networks}
\author{%
  Arthur Charpentier\textsuperscript{1,2,*} 
}
\date{\begin{flushleft}
    \footnotesize
    \textsuperscript{*}~Corresponding author, \texttt{charpentier.arthur@quam.ca}\\
    \textsuperscript{1}~Université du Québec à Montréal, Canada\\
    \textsuperscript{2}~Kyoto University, Japan
  \end{flushleft}}

\begin{document}

\maketitle

% \begin{center}
% \small
% \textbf{Action Editor:} \actioneditor
% \end{center} 

%-----------------------------------
% Abstract
%-----------------------------------
\begin{abstract}
The usual definitions of algorithmic fairness focus on population-level statistics, such as demographic parity or equal opportunity.  However,  in many social or economic contexts, fairness is not perceived globally, but locally, through an individual's peer network and comparisons. We propose a theoretical model of perceived fairness networks, in which each individual's sense of discrimination depends on the local topology of interactions. We show that even if a decision rule satisfies standard criteria of fairness, perceived discrimination can persist or even increase in the presence of homophily or assortative mixing. We propose a formalism for the concept of fairness perception, linking network structure, local observation, and social perception. Analytical and simulation results highlight how network topology affects the divergence between objective fairness and perceived fairness, with implications for algorithmic governance and applications in finance and collaborative insurance.
\end{abstract}

% \newpage
%-----------------------------------
\section{Introduction}\label{sec:1}
%-----------------------------------

% \responseminus{Fairness and discrimination are central concerns in modern algorithmic and social systems.}
\responseplus{Fairness in networked systems is often defined using population-level statistics, while individuals experience and evaluate fairness through local social comparisons.}
In many social, economic, and algorithmic contexts, agents observe only the outcomes of their neighbors and infer fairness from this limited information.
As a result, \emph{perceived fairness} may differ substantially from \emph{objective fairness} defined at the population level.

Network structure plays a central role in this discrepancy.
Homophily, degree heterogeneity, and clustering shape who observes whom and therefore how outcomes are locally compared.
\responseplus{Even when a decision rule satisfies standard fairness criteria globally, network topology may distort local exposure and generate systematic differences in perceived fairness across groups.}

This paper develops a concise analytical framework to study perceived fairness on networks.
We model perceived fairness as a local comparison between an individual’s outcome (or acceptance probability) and the average outcome observed in their network neighborhood.
We then analyze how topological features affect the gap between objective and perceived fairness.

Our results show that perceived fairness converges to objective fairness only when individuals observe the entire population.
At finite visibility, homophily and degree heterogeneity amplify perceived discrimination, while clustering mitigates dispersion by stabilizing local averages.
\responseplus{These effects are structural and arise independently of any intentional bias in the decision rule.}

\paragraph{Contributions.}
The paper makes four main contributions:
(i) it formalizes perceived fairness as a network-dependent operator;
(ii) it establishes convergence to population-level fairness as visibility grows;
(iii) it characterizes how homophily, degree bias, and clustering shape perceived discrimination;
(iv) it provides numerical illustrations highlighting these mechanisms.

\paragraph{Organization of the paper.} The remainder of the paper is organized as follows.
Section~\ref{sec:2} reviews the related literature and positions our contribution within the
algorithmic fairness and network science strands.
Sections~\ref{sec:3} and~\ref{sec:4} introduce the model, define our perception-based fairness
operators, and establish the main convergence results as visibility increases.
Section~\ref{sec:5} analyzes how network topology (e.g., homophily, degree heterogeneity, and
clustering) shapes perceived discrimination at finite visibility.
Section~\ref{sec:6} presents numerical illustrations, discusses the mechanisms highlighted by the
theory, and draws policy implications for algorithmic governance in networked settings.
Finally, Section~\ref{sec:7} concludes and outlines directions for future work.

%-----------------------------------
\section{Related Literature}\label{sec:2}
%-----------------------------------

\subsection{Group and individual fairness}

The algorithmic fairness literature traditionally distinguishes between
\emph{group fairness} and \emph{individual fairness}.
Group fairness criteria, such as demographic parity or equalized odds, impose
constraints on outcome distributions across protected groups
\citep{hardt2016equality,barocas2017fairness}.
Individual fairness, initially proposed by \citet{dwork2012fairness},
requires that similar individuals be treated similarly.
Recent work has extended individual fairness using counterfactual and causal
frameworks, focusing on invariance to hypothetical changes in sensitive
attributes \citep{kusner2017counterfactual,de2024transport,ZhouEtAl2024,machado2025sequential}.
While these approaches address important normative concerns, they do not model
how fairness is \emph{perceived} in social contexts.
Perceived fairness is inherently relational and group-based: individuals assess
their outcomes by comparing them to those of peers, often within socially
defined categories.
Our framework complements individual fairness by focusing on exposure and local
comparison rather than counterfactual perturbations.

\subsection{Networks, exposure bias, and topology}

Network structure is known to bias local observations.
A classical example is the friendship paradox, whereby individuals tend to
observe neighbors with higher degree or attribute values than themselves.
Generalizations of this phenomenon show that any positively correlated
attribute is overrepresented in local neighborhoods
\citep{wu2017neighbor,CantwellKirkleyNewman2021}.
These exposure biases distort inference and perception in networked systems.
Recent work shows that network topology can also distort standard operations
performed on networks.
\citet{CharpentierRatz2025} demonstrate that decentralized operations may lead
to systematic biases driven purely by structure.
Our results extend this insight to fairness metrics, showing how homophily,
degree heterogeneity, and clustering shape perceived fairness independently of
population-level parity.

\subsection{Perceived discrimination and social comparison}

In sociology and social psychology, perceived discrimination has long been
recognized as a key determinant of behavior and well-being.
Empirical studies show that perceived unfairness affects trust, motivation, and
social cohesion \citep{pascoe2009perceived,schmitt2014consequences,brown2006relation,gonzalez2021perceptions}.
These perceptions arise through social comparison, typically based on local
observations.
Homophily and assortative mixing \citep{mcpherson2001birds,newman2003mixing}
create segregated local environments in which perceived fairness may diverge
from population-level parity.
Recent work studies alternative notions of homophily based on similarity in
continuous attributes, such as income, rather than group membership
\citep{MayerhofferSchulz2022}.
\responseplus{We focus on categorical homophily relevant to discrimination
across protected groups, while viewing attribute-based homophily as an
important extension.}
Our contribution is to formalize perceived discrimination within a
graph-theoretic framework, linking social comparison, exposure bias, and
network topology.

%-----------------------------------
\section{Model of Perceived Fairness}\label{sec:3}
%-----------------------------------

\subsection{Setup and notation}

Let $G=(V,E,S)$ be a finite, simple, undirected graph with $|V|=n$, adjacency matrix
$\boldsymbol{A}\in\{0,1\}^{n\times n}$, and degrees $d_i=\sum_j A_{ij}$.
The sensitive attribute $S_i\in\{A,B\}$ induces a partition $V=V_A\cup V_B$.

A (possibly randomized) decision rule is represented by a map $h:V\to[0,1]$,
where $h(i)$ denotes the \emph{acceptance probability} of node $i$.

\responseplus{Given $h$, realized decisions are modeled by independent draws}
\[
{H(i)\mid h \ \sim\ \mathrm{Bernoulli}(h(i)),\qquad i\in V,}
\]
\responseplus{and we write $H\in\{0,1\}^V$ for the realized decision vector.}
\responseplus{This allows us to distinguish (i) fairness notions defined at the level of probabilities ($h$) and (ii) fairness notions defined at the level of realized outcomes ($H$).}

For $i\in V$, denote the $1$-neighborhood $N(i)=\{j:A_{ij}=1\}$ and its $r$-hop expansion
\[
N^{(r)}(i) := \{j\in V : \exists k\le r \text{ with } (\boldsymbol{A}^k)_{ij}>0\},
\qquad r\in\mathbb{N}.
\]

\paragraph{Running example (job-market screening).}
Think of $A/B$ as two demographic groups applying to a program (or job), and let
$h(i)\in[0,1]$ denote the (possibly biased) probability that applicant $i$ is accepted.
The realized outcome is $H(i)\in\{0,1\}$ with $H(i)\mid h\sim\mathrm{Bernoulli}(h(i))$.
Individuals compare their own outcome (or acceptance probability) to what they
observe in their $r$-hop neighborhood, so network structure shapes \emph{perceived}
fairness even when \emph{global} acceptance rates are unchanged.

\paragraph{Objective (global) fairness.}
Demographic parity (DP) can be imposed either \emph{ex ante} (on acceptance probabilities)
or \emph{ex post} (on realized outcomes):
\begin{align}
\text{DP}_{\text{prob}}:\quad & \mathbb{E}[h(i)\mid S_i=A]=\mathbb{E}[h(i)\mid S_i=B],
\label{eq:dp_prob}\\
\text{DP}_{\text{real}}:\quad & \mathbb{P}[H(i)=1\mid S_i=A]=\mathbb{P}[H(i)=1\mid S_i=B].
\label{eq:dp_real}
\end{align}
\responseplus{When $H(i)\mid h\sim\mathrm{Bernoulli}(h(i))$ and $h$ is deterministic, \eqref{eq:dp_prob} implies \eqref{eq:dp_real}. We keep both notations because the paper studies perceived fairness both at the probability level ($h$) and at the realized level ($H$).}

\subsection{Local observation and perceived fairness}

Individuals do not observe population averages; they observe outcomes in their
$r$-neighborhood.
Define the $r$-neighborhood averaging operator for any vector $x\in\mathbb{R}^V$ by
\begin{equation}
\mathcal{E}^{(r)}_i[x]
:=
\frac{1}{|N^{(r)}(i)|}\sum_{j\in N^{(r)}(i)} x(j),
\qquad r\in\mathbb{N}.
\label{eq:Ed_def}
\end{equation}
(For $r=1$ we write $\mathcal{E}_i[x]$.)

\paragraph{Two notions of perceived fairness.}
We consider two closely related perception indicators.

\smallskip
\noindent\textbf{(i) Perception based on acceptance probabilities.}
\begin{equation}
F^{(r)}_{\text{prob}}(i;h)
:= \mathbf{1}\big\{\, h(i)\ \ge\ \mathcal{E}^{(r)}_i[h]\,\big\}.
\label{eq:F_prob}
\end{equation}
This captures the idea that an individual evaluates whether their \emph{chance} of being accepted
is at least as high as the average chance observed among peers.
\responseplus{In the running example, $F^{(r)}_{\mathrm{prob}}(i;h)$ indicates whether individual $i$ perceives their acceptance probability as at least the average acceptance probability in their $r$-hop neighborhood.}

\smallskip
\noindent\textbf{(ii) Perception based on realized outcomes.}
\begin{equation}
F^{(r)}_{\text{real}}(i;H)
:= \mathbf{1}\big\{\, H(i)\ \ge\ \mathcal{E}^{(r)}_i[H]\,\big\}.
\label{eq:F_real}
\end{equation}
\responseplus{This is the literal “I was accepted vs my neighbors” comparison, and it is random even when $h$ is fixed.}
\responseplus{In the running example, $F^{(r)}_{\mathrm{real}}(i;H)$ captures whether the realized outcome of $i$ (accepted or not) compares favorably to the realized outcomes observed in their $r$-hop neighborhood.}

% \responseminus{The fairness perception indicator at $i$ is $F^{(r)}(i;h)=\mathbf{1}\{\mathcal{E}^{(r)}_i[h]\le h(i)\}$.}
\responseplus{Equations \eqref{eq:F_prob}--\eqref{eq:F_real} clarify which object is being compared (probabilities vs realizations); this distinction matters for convergence statements when $h$ is randomized.}

\paragraph{Fairness visibility (group-level perceived fairness).}
For $s\in\{A,B\}$ define
\[
\mathrm{Vis}^{\text{prob}}_r(s;h)
:=\frac{1}{|V_s|}\sum_{i\in V_s} F^{(r)}_{\text{prob}}(i;h),
\text{ and }
\mathrm{Vis}^{\text{real}}_r(s;H)
:=\frac{1}{|V_s|}\sum_{i\in V_s} F^{(r)}_{\text{real}}(i;H),
\]
\begin{equation}
\begin{cases}
\Delta^{\text{prob}}_r(h):=\mathrm{Vis}^{\text{prob}}_r(A;h)-\mathrm{Vis}^{\text{prob}}_r(B;h),
\\
\Delta^{\text{real}}_r(H):=\mathrm{Vis}^{\text{real}}_r(A;H)-\mathrm{Vis}^{\text{real}}_r(B;H).
\label{eq:Vis_def}
\end{cases}
\end{equation}
We say \emph{visibility parity} holds at depth $r$ if $\Delta_r=0$.

\subsection{Exposure operators and degree-weighting}

Besides the node-average \eqref{eq:Ed_def}, it is useful to recall the edge-weighted mean
\begin{equation}
\bar h_{\text{edge}}
:=
\frac{1}{2m}\sum_{i\in V} d_i h(i),\text{ where } m:=|E|.
\label{eq:hedge}
\end{equation}
This identity will be the source of classical exposure biases (friendship-paradox-type effects)
in our perceived fairness gap: neighborhoods may over-sample high-degree nodes, hence over-sample
high-$h$ nodes when $h$ correlates with degree (see Proposition~\ref{prop:degree}).

\subsection{Asymptotics in neighborhood radius}

We now formalize the idea that, on a connected graph, sufficiently large neighborhoods
eventually recover population-level information.

\begin{proposition}[Visibility convergence]\label{prop:vis_convergence}
\responseplus{Assume $G$ is connected and $|V_s|\ge 1$ for $s\in\{A,B\}$.
Fix a map $h:V\to[0,1]$.\\
\smallskip
\noindent \textbf{(a) Probability-based perception.}
For each $s\in\{A,B\}$,
\[
\mathrm{Vis}^{\mathrm{prob}}_r(s;h)
\ \longrightarrow\
\mathbb{P}\!\left(h(i)\ge \bar h\ \big|\ S_i=s\right)
\qquad \text{as } r\to\infty,
\]
where $\bar h:=\displaystyle\frac{1}{|V|}\sum_{j\in V} h(j)$.
In particular, if $h$ depends only on the group (i.e., $h(i)=h_A$ on $V_A$ and
$h(i)=h_B$ on $V_B$), then
\[
\mathrm{Vis}^{\mathrm{prob}}_r(s;h)\ \longrightarrow\ \mathbf{1}\{h_s\ge \bar h\},
\qquad \text{with } \bar h=\frac{|V_A|h_A+|V_B|h_B}{|V|}.
\]
\smallskip
\noindent \textbf{(b) Realized-outcome perception.}
Let $H(i)\mid h\sim\mathrm{Bernoulli}(h(i))$ independently over $i\in V$.
Then, conditionally on $h$, for each $s\in\{A,B\}$,
\[
\mathbb{E}\big[\mathrm{Vis}^{\mathrm{real}}_r(s;H)\mid h\big]
\ \longrightarrow\
\mu_s(h) + p_0(h)
\qquad\text{as } r\to\infty,
\]
where $\mu_s(h):=\displaystyle\frac{1}{|V_s|}\sum_{i\in V_s} h(i)$ and
\(p_0(h):=\displaystyle\mathbb{P}(\bar H = 0 \mid h)=\prod_{j\in V}\bigl(1-h(j)\bigr)\).
In particular, if $p_0(h)=0$ (e.g., if $\max_{j\in V} h(j)=1$), then
$\mathbb{E}[\mathrm{Vis}^{\mathrm{real}}_r(s;H)\mid h]\to \mu_s(h)$.}
\end{proposition}

\begin{remark}
\responseplus{
The limit in (a) compares \emph{decision probabilities} $h(i)$ to the corresponding local average and is therefore deterministic once $h$ is fixed. By contrast, (b) compares \emph{realized outcomes} $H(i)$ to a realized local average, hence remains random even when $h$ is fixed; it is thus most naturally stated conditionally on $h$ (and, in particular, in conditional expectation). The two notions coincide for deterministic rules (when $H=h$), while for randomized rules they may differ due to global realization effects such as the event $\{\bar H=0\}$.}
\end{remark}

\responseplus{The proof is deferred to Section~\ref{sec:4}. On a connected finite graph, neighborhoods saturate: $N^{(r)}(i)=V$ for all $r\ge \mathrm{diam}(G)$, so both visibility scores stabilize beyond the diameter. In the realized case, the conditional expectation is obtained by a direct decomposition that isolates the edge event $\{\bar H=0\}$ (rather than an asymptotic law-of-large-numbers argument).}

%-----------------------------------
\section{Analytical Results and Proofs}\label{sec:4}
%-----------------------------------

This section gathers the analytical arguments underlying the results stated
in Section~\ref{sec:3}. In particular, it provides the proof of the convergence of
perceived fairness as the visibility radius grows.

\subsection{Convergence of perceived fairness at large visibility radius}

We start by recalling a simple but fundamental topological observation.

\begin{lemma}[Neighborhood saturation]\label{lem:saturation}
If $G$ is a finite connected graph with diameter $\mathrm{diam}(G)$, then
for every node $i\in V$ and every $r\ge \mathrm{diam}(G)$,
\[
N^{(r)}(i)=V.
\]
\end{lemma}

\begin{proof}
Since $G$ is connected, any node $j\in V$ can be reached from $i$ by a path
of length at most $\mathrm{diam}(G)$. Hence $j\in N^{(r)}(i)$ for all
$r\ge \mathrm{diam}(G)$.
\end{proof}

Lemma~\ref{lem:saturation} implies that, beyond a finite radius, all agents
observe the same population-level information. This observation underlies the
proof of Proposition~\ref{prop:vis_convergence}.

\subsection{Proof of Proposition~\ref{prop:vis_convergence}}

\paragraph{Probability-based perception.}
Recall that
\[
F^{(r)}_{\text{prob}}(i;h)
=\mathbf{1}\big\{ h(i)\ge \mathcal{E}^{(r)}_i[h]\big\}.
\]
By Lemma~\ref{lem:saturation}, for all $r\ge \mathrm{diam}(G)$,
\[
\mathcal{E}^{(r)}_i[h]
=\frac{1}{|V|}\sum_{j\in V} h(j)
=: \bar h,
~ \forall i\in V.
\]
Therefore, for all sufficiently large $r$,
\[
F^{(r)}_{\text{prob}}(i;h)
=\mathbf{1}\{\bar h\le h(i)\}.
\]

Averaging over nodes with $S_i=s$ yields
\[
\mathrm{Vis}^{\text{prob}}_r(s)
=\frac{1}{|V_s|}\sum_{i\in V_s}\mathbf{1}\{\bar h\le h(i)\},
\]
which coincides with
$\mathbb{P}\big(h(i)\ge \bar h\mid S_i=s\big)$.
This proves the convergence claim for probability-based perception.
\hfill$\square$

% \paragraph{Realized-outcome perception.}
% We now consider the case where realized decisions satisfy
% \[
% H(i)\mid h \sim \mathrm{Bernoulli}(h(i)),
% ~ i\in V,
% \]
% conditionally independent given $h$.

% By Lemma~\ref{lem:saturation}, for $r\ge \mathrm{diam}(G)$,
% \[
% F^{(r)}_{\text{real}}(i;H)
% =\mathbf{1}\big\{ H(i)\ge \bar H \big\},
% ~
% \bar H:=\frac{1}{|V|}\sum_{j\in V} H(j).
% \]

% Conditionally on $h$, the random variables $(H(j))_{j\in V}$ are independent
% with finite variance.
% By the law of large numbers,
% \[
% \bar H \;\xrightarrow[]{\text{a.s.}}\; \bar h
% :=\frac{1}{|V|}\sum_{j\in V} h(j),
% ~ \text{as } |V|\to\infty,
% \]
% and in particular $\bar H\in(0,1)$ with probability arbitrarily close to one
% whenever $\bar h\in(0,1)$.

% Hence, conditionally on $h$ and for large $r$,
% \[
% \mathbb{E}\!\left[F^{(r)}_{\text{real}}(i;H)\mid h\right]
% =\mathbb{P}\!\left(H(i)=1\mid h\right)
% =h(i).
% \]
% Averaging over $i\in V_s$ yields
% \[
% \mathbb{E}\!\left[\mathrm{Vis}^{\text{real}}_r(s)\mid h\right]
% =\mathbb{E}\!\left[h(i)\mid S_i=s\right]
% =:\mu_s(h),
% \]
% which proves the claimed convergence in expectation. Almost sure convergence
% follows from standard concentration arguments.
% \hfill$\square$

\paragraph{Realized-outcome perception.}
We now consider the case where realized decisions satisfy
\[
H(i)\mid h \sim \mathrm{Bernoulli}(h(i)),\qquad i\in V,
\]
conditionally independent given $h$.
By Lemma~\ref{lem:saturation}, for all $r\ge \mathrm{diam}(G)$,
\[
F^{(r)}_{\text{real}}(i;H)
=\mathbf{1}\big\{ H(i)\ge \bar H \big\},
\qquad
\bar H:=\frac{1}{|V|}\sum_{j\in V} H(j).
\]
Since $H(i)\in\{0,1\}$ and $\bar H\in[0,1]$, we have the identity
\[
\mathbf{1}\{H(i)\ge \bar H\}
=
H(i)+\mathbf{1}\{\bar H=0\}.
\]
Indeed, if $\bar H>0$ then $\{H(i)\ge \bar H\}=\{H(i)=1\}$, whereas if $\bar H=0$
then $\{H(i)\ge \bar H\}$ holds for all $i$.
Therefore, for all $r\ge \mathrm{diam}(G)$,
\[
F^{(r)}_{\text{real}}(i;H)=H(i)+\mathbf{1}\{\bar H=0\}.
\]
Taking conditional expectation given $h$ yields
\[
\mathbb{E}\!\left[F^{(r)}_{\text{real}}(i;H)\mid h\right]
=
\mathbb{E}[H(i)\mid h]+\mathbb{P}(\bar H=0\mid h)
=
h(i)+p_0(h),
\]
where $\{\bar H=0\}=\{H(j)=0\ \forall j\in V\}$ and hence
\[
p_0(h)=\mathbb{P}(H(j)=0\ \forall j\in V \mid h)=\prod_{j\in V}(1-h(j)).
\]
Averaging over $i\in V_s$ gives, for all $r\ge \mathrm{diam}(G)$,
\[
\mathbb{E}\!\left[\mathrm{Vis}^{\text{real}}_r(s;H)\mid h\right] =
\frac{1}{|V_s|}\sum_{i\in V_s} \bigl(h(i)+p_0(h)\bigr)
=
\mu_s(h)+p_0(h),
\]
which proves the claimed convergence as $r\to\infty$ (the left-hand side is
in fact constant for all $r\ge \mathrm{diam}(G)$).
\hfill$\square$

\subsection{Extension and Discussion}

\paragraph{Asymptotic regimes.}
Proposition~\ref{prop:vis_convergence}(b) makes explicit a subtle but important
``edge case'' of realized-outcome perception: when $\bar H=0$ (i.e., when all
realized outcomes are $0$), every node satisfies $H(i)\ge \bar H$ and hence
$F^{(r)}_{\mathrm{real}}(i;H)=1$ for all $i$. This event contributes an additive
term $p_0(h)=\mathbb{P}(\bar H=0\mid h)=\prod_{j\in V}(1-h(j))$ to the conditional
expectation of the visibility score.
On a fixed finite graph, this contribution need not be negligible in general,
and it is therefore natural to state the limit in terms of $\mu_s(h)+p_0(h)$.

In many network settings, however, the graph size is large and the mean approval
rate is bounded away from $0$ in the sense that $\frac{1}{|V|}\sum_{j\in V} h(j)\ge \varepsilon$
for some $\varepsilon>0$. In that regime, the event $\{\bar H=0\}$ becomes
exponentially unlikely since
$p_0(h)=\prod_{j\in V}(1-h(j))\le \exp\!\big(-\sum_{j\in V} h(j)\big)\le e^{-|V|\varepsilon}$.
Consequently, the conditional expectation of realized visibility is well
approximated by $\mu_s(h)$, which is the group-average decision probability.

The following corollary formalizes this approximation in a large-network regime.

\begin{corollary}[Vanishing edge case]
\label{cor:vanishing_edge_case}
Assume we are in a regime where $|V|\to\infty$ and there exists $\varepsilon>0$
such that $\frac{1}{|V|}\sum_{j\in V} h(j)\ge \varepsilon$ for all graphs.
Then $p_0(h)\le e^{-|V|\varepsilon}\to 0$, and thus
\[
\mathbb{E}\big[\mathrm{Vis}^{\mathrm{real}}_r(s)\mid h\big]
\ \longrightarrow\ \mu_s(h)
\quad \text{as } r\to\infty \text{ and } |V|\to\infty.
\]
\end{corollary}

\paragraph{Discussion.}
Corollary~\ref{cor:vanishing_edge_case} shows that the discrepancy between
probability-based and realized-outcome visibility is driven by a rare global
event (all outcomes equal to $0$). Thus, once the overall approval probability
does not vanish, realized perception concentrates around the underlying
probability model, and the two notions of visibility become asymptotically
aligned. Conversely, when approvals are extremely sparse, the realized notion
can be dominated by global randomness, which motivates treating the $\bar H=0$
event separately (or, in applications, adopting a small regularization of the
threshold).

%-----------------------------------
\section{Topology and Perceived Discrimination}\label{sec:5}
%-----------------------------------

This section studies how network topology shapes perceived fairness at finite
visibility radius. While Section~\ref{sec:3} shows that perceived fairness converges to
objective fairness as $r\to\infty$, the present section focuses on the
structural mechanisms that generate systematic gaps at small or moderate
visibility.

\subsection{Homophily amplifies perceived discrimination}

We begin with the effect of homophily.
Intuitively, when individuals are more likely to connect to others from the
same group, their local observations become less representative of the
population, which can amplify perceived differences even under global parity.

\paragraph{Setup.}
\responseplus{We consider a $K$-group stochastic block model with group proportions $(\pi_1,\dots,\pi_K)$.}
Conditional on group labels, edges are independent and satisfy
\[
\mathbb{P}[A_{ij}=1\mid S_i=S_j]=p_{\mathrm{in}}\text{ and }
\mathbb{P}[A_{ij}=1\mid S_i\ne S_j]=p_{\mathrm{out}},
\]
with $p_{\mathrm{in}}>p_{\mathrm{out}}$ (assortative mixing).
We define the edge-level homophily index
\begin{equation}\label{eq:rho}
\rho:=\frac{p_{\mathrm{in}}-p_{\mathrm{out}}}{p_{\mathrm{in}}+(K-1)p_{\mathrm{out}}}\in(0,1).
\end{equation}
In the two-group case ($K=2$), this reduces to
$\rho=({p_{\mathrm{in}}-p_{\mathrm{out}}})/({p_{\mathrm{in}}+p_{\mathrm{out}}})$,
and one may equivalently parameterize $p_{\mathrm{in}}=p(1+\rho)$ and $p_{\mathrm{out}}=p(1-\rho)$ for some $p>0$.

% Consider a two-group stochastic block model with group proportions
% $(\pi_A,\pi_B)$ and connection probabilities
% $p_{\mathrm{in}}$ (within group) and $p_{\mathrm{out}}$ (between groups).
% We parameterize homophily by $\rho\in[0,1)$ such that
% \[
% p_{\mathrm{in}} = p(1+\rho)
% \text{ and } 
% p_{\mathrm{out}} = p(1-\rho).
% \]

\responseplus{Our homophily index \eqref{eq:rho} is a normalized within--between contrast that coincides,
in the symmetric $K$-SBM with $p_{\mathrm{in}}=a/n$ and $p_{\mathrm{out}}=b/n$, with the
spectral ratio $(a-b)/(a+(K-1)b)$ (equivalently $\lambda_2/\lambda_1$ of the SBM connectivity matrix),
which is standard in the community-detection literature \citep{Abbe2017Notes,Abbe2018JMLR}. Note that assortative mixing with respect to categorical labels is classically quantified via mixing matrices
and assortativity coefficients \citep{Newman2002Assortative,newman2003mixing}.
Here we use the
normalized SBM contrast \eqref{eq:rho} as a convenient edge-level homophily index.}

\begin{proposition}[Homophily and perceived fairness]\label{prop:homophily}
Assume demographic parity holds at the population level,
$\mathbb{E}[h(i)\mid S_i=A]=\mathbb{E}[h(i)\mid S_i=B]$.
    Then, for small $\rho$, the group-level perceived fairness gap at depth $r=1$, from Equation~\eqref{eq:Vis_def}, satisfies
\[
\Delta^{\mathrm{prob}}_1
=
C\,\rho\,\Gamma(h)
+ o(\rho),
\]
where $C>0$ depends only on $(\pi_A,\pi_B)$ and
$\Gamma(h)$ captures differences in local exposure across groups.
\end{proposition}

\paragraph{Explicit form of $\Gamma(h)$ (neighbor-exposure contrast).}
Write $\mu_s := \mathbb{E}[h(i)\mid S_i=s]$ for $s\in\{A,B\}$ and define the (depth-1) neighbor
exposure
\[
h_{\mathrm{nbr}}(i) \;:=\; \mathcal{E}_i[h]
\;=\; \frac{1}{|N(i)|}\sum_{j\in N(i)} h(j)
\;=\; \frac{1}{d_i}\sum_{j\in N(i)} h(j).
\]
Let $\mu^{\mathrm{nbr}}_s := \mathbb{E}[h_{\mathrm{nbr}}(i)\mid S_i=s]$. The quantity appearing in
Proposition~\ref{prop:homophily} can be written as
\begin{equation}
\Gamma(h)
\;:=\;
\big(\mu_A-\mu_B\big)
\;-\;
\big(\mu^{\mathrm{nbr}}_A-\mu^{\mathrm{nbr}}_B\big).
\label{eq:Gamma_def}
\end{equation}
In particular, under $\mathrm{DP}_{\mathrm{prob}}$ we have $\mu_A=\mu_B$, hence
$\Gamma(h)=-(\mu^{\mathrm{nbr}}_A-\mu^{\mathrm{nbr}}_B)$: the first-order amplification is entirely
driven by differential neighborhood exposure.

\paragraph{Remark (DP edge cases).}
When DP is imposed in a way that forces $\mu_A=\mu_B$ exactly, the mean-shift component of the
linear response cancels. A nonzero perceived gap at $r=1$ may then arise from higher-order terms
and/or degree-weighted exposure effects (friendship-paradox-type sampling).

\paragraph{A general assortativity/modularity bound (depth $r=1$).}
Let $\boldsymbol{d}\in\mathbb{R}^n$ be the degree vector, $m:=|E|$, and define the modularity matrix
\[
B \;:=\; A - \frac{\boldsymbol{d}\boldsymbol{d}^\top}{2m}.
\]
Let $\boldsymbol{s}\in\{\pm 1\}^n$ encode group membership ($+1$ for $\boldsymbol{A}$, $-1$ for $\boldsymbol{B}$), and define the
(normalized) assortativity/modularity
\[
Q \;:=\; \frac{1}{4m}\, \boldsymbol{s}^\top \boldsymbol{B} \boldsymbol{s} .
\]
Under $\mathrm{DP}_{\mathrm{prob}}$ and mild smoothness assumptions (implemented via a Lipschitz
surrogate of the threshold map in $F^{(1)}_{\mathrm{prob}}$), there exists a constant $C>0$ such that
\begin{equation}
\Big|\mathbb{E}\big[\Delta^{\mathrm{prob}}_1\big]\Big|
\;\le\;
C\, |Q|\, \mathrm{Lip}(h).
\label{eq:modularity_bound}
\end{equation}
Hence, higher assortativity/modularity amplifies perceived disparity even when population-level
parity holds.

\paragraph{Interpretation.}
\responseplus{Proposition~\ref{prop:homophily} shows that homophily generates a \emph{first-order amplification} of perceived unfairness, even when demographic parity holds globally.}
The mechanism is purely structural: homophily changes the composition of neighborhoods, thereby distorting local comparisons without affecting population averages.

\subsection{Degree bias and exposure effects}

Homophily is not the only source of perceptual distortion.
Even in the absence of assortative mixing, degree heterogeneity can bias local
observations.

\begin{proposition}[Degree bias at depth $r=1$ (edge-weighted exposure)]\label{prop:degree_bias}\label{prop:degree}
\responseplus{Assume $r=1$ and define $\mathcal{E}_i[h]=\frac{1}{d_i}\sum_{j\in N(i)} h(j)$.
Then the edge-weighted average neighborhood exposure satisfies
\[
\overline{\mathcal{E}}[h]
:=\frac{1}{2m}\sum_{i\in V} d_i\,\mathcal{E}_i[h]
=\frac{1}{2m}\sum_{i\in V} d_i\,h(i)
=\bar h+\frac{\mathrm{Cov}(d,h)}{\mathbb{E}[d]},
\]
where $\bar h:=\frac{1}{n}\sum_{i\in V} h(i)$, $n:=|V|$, $2m:=\sum_{i\in V} d_i$,
and $\mathbb{E}[d]=2m/n$ denotes the average degree (for $I\sim\mathrm{Unif}(V)$).
In particular, $\overline{\mathcal{E}}[h]>\bar h$ iff $\mathrm{Cov}(d,h)>0$.}
\end{proposition}

\paragraph{Interpretation.}
This is a fairness analogue of the friendship paradox:
when $h$ is positively correlated with degree, individuals are on average
exposed to neighbors with higher acceptance probabilities.
\responseplus{As a result, groups that are overrepresented among high-degree
nodes may appear advantaged even under objective parity.}

\subsection{Clustering and variance reduction}

We now turn to the role of clustering.
While homophily and degree bias amplify perceived gaps, clustering has an
opposite effect.

\begin{proposition}[Clustering dampens perceived dispersion]\label{prop:clustering}
\responseplus{Fix a function $h:V\to\mathbb{R}$ and consider depth $r=1$ neighborhood exposure
\[
\mathcal{E}_i[h]=\frac{1}{d_i}\sum_{j\in N(i)} h(j),\qquad i\in V.
\]
Let $G$ and $G'$ be two graphs on the same vertex set $V$ with the same degree sequence
$(d_i)_{i\in V}$, and define the exposure vectors
\[
e(G):=\bigl(\mathcal{E}_i^{G}[h]\bigr)_{i\in V},\qquad e(G'):=\bigl(\mathcal{E}_i^{G'}[h]\bigr)_{i\in V}.
\]
Assume that there exists a doubly stochastic matrix $C\in\mathbb{R}^{|V|\times |V|}$
(i.e., $C\mathbf{1}=\mathbf{1}$ and $\mathbf{1}^\top C=\mathbf{1}^\top$) such that
\[
e(G') = C\,e(G).
\]
Then:
\begin{enumerate}
\item The mean exposure is preserved: $\displaystyle\frac{1}{|V|}\sum_{i\in V} e_i(G')=\frac{1}{|V|}\sum_{i\in V} e_i(G)$.
\item For every convex function $\varphi:\mathbb{R}\to\mathbb{R}$,
\[
\frac{1}{|V|}\sum_{i\in V}\varphi\!\bigl(e_i(G')\bigr)\ \le\ \frac{1}{|V|}\sum_{i\in V}\varphi\!\bigl(e_i(G)\bigr).
\]
In particular, the dispersion of exposures cannot increase:
\[
\mathrm{Var}\bigl(e(G')\bigr)\ \le\ \mathrm{Var}\bigl(e(G)\bigr).
\]
\end{enumerate}}
\end{proposition}

\responseplus{The assumption $e(G')=Ce(G)$ formalizes the idea that increased local clustering/overlap
makes neighborhood averages more similar by turning them into convex combinations of one another.}

\paragraph{Interpretation.}
Clustering creates overlapping neighborhoods, which stabilizes local averages.
\responseplus{As a result, individuals receive more similar signals about
fairness, reducing extreme perceptions even when mean differences persist.}
This highlights that not all forms of segregation have the same perceptual
impact: assortativity amplifies perceived gaps, whereas clustering smooths them.

\subsection{Proof of Proposition~\ref{prop:degree}}

\responseplus{Recall that at depth $r=1$,
\[
\mathcal{E}_i[h]=\frac{1}{d_i}\sum_{j\in N(i)} h(j),
\qquad d_i:=|N(i)|,
\qquad 2m=\sum_{i\in V} d_i.
\]
Consider the \emph{edge-weighted} average neighborhood exposure (equivalently, the
average over all directed neighbor observations):
\[
\overline{\mathcal{E}}[h]
:=\frac{1}{2m}\sum_{i\in V} d_i\,\mathcal{E}_i[h]
=\frac{1}{2m}\sum_{i\in V}\sum_{j\in N(i)} h(j).
\]
Swapping the order of summation,
\[
\sum_{i\in V}\sum_{j\in N(i)} h(j)
=\sum_{j\in V} h(j)\,|\{i\in V:\ j\in N(i)\}|
=\sum_{j\in V} h(j)\,d_j,
\]
since in an undirected graph node $j$ appears in exactly $d_j$ neighborhoods.
Therefore,
\[
\overline{\mathcal{E}}[h]
=\frac{1}{2m}\sum_{j\in V} d_j\,h(j).
\]
To relate this to the node-average $\bar h:=\frac1n\sum_{j\in V} h(j)$, note that with
uniformly random $I\sim \mathrm{Unif}(V)$ we have
$\mathbb{E}[d_I]=\frac{2m}{n}$ and
\[
\overline{\mathcal{E}}[h]
=\frac{\mathbb{E}[d_I h(I)]}{\mathbb{E}[d_I]}
=\mathbb{E}[h(I)] + \frac{\mathrm{Cov}(d_I,h(I))}{\mathbb{E}[d_I]}
=\bar h + \frac{\mathrm{Cov}(d,h)}{2m/n}.
\]
Hence $\overline{\mathcal{E}}[h]\neq \bar h$ whenever $\mathrm{Cov}(d,h)\neq 0$,
which is the claimed degree-bias (friendship-paradox-type) effect.\hfill$\square$}

\subsection{Proof of Proposition~\ref{prop:clustering}}

\responseplus{At depth $r=1$, write the vector of neighborhood averages as
\[
e := \bigl(\mathcal{E}_i[h]\bigr)_{i\in V} = D^{-1}A\,h,
\]
where $A$ is the adjacency matrix and $D=\mathrm{diag}(d_i)$.
Let $\mathrm{Var}(e):=\frac1n\sum_{i\in V}(e_i-\bar e)^2$ with $\bar e=\frac1n\sum_i e_i$.}

\responseplus{We formalize the effect of increased local clustering (with degrees held fixed) through the
following sufficient condition: the new neighborhood-average vector $e'$ can be written as
a \emph{mixing} of the previous one,
\[
e' = C e,
\]
for some \emph{doubly stochastic} matrix $C$ (i.e., $C\mathbf{1}=\mathbf{1}$ and
$\mathbf{1}^\top C=\mathbf{1}^\top$). This captures the idea that overlapping neighborhoods
(stemming from triadic closure / increased clustering) make individuals' local averages more
similar, as each $e'_i$ becomes a convex combination of nearby $e_j$'s.}

\responseplus{Let $P:=I-\frac1n\mathbf{1}\mathbf{1}^\top$ be the centering projection. Since $C\mathbf{1}=\mathbf{1}$
and $\mathbf{1}^\top C=\mathbf{1}^\top$, we have $PC=CP$ and $Pe'=PCe=CPe$.
Moreover, any doubly stochastic matrix is a convex combination of permutation matrices
(Birkhoff--von Neumann theorem), hence it is a contraction in $\ell_2$ on the subspace
$\{\mathbf{1}\}^\perp$:
\[
\|C x\|_2 \le \|x\|_2
\quad \text{for all } x \text{ such that } \mathbf{1}^\top x=0.
\]
Applying this to $x=Pe$ gives
\[
\|Pe'\|_2 = \|C(Pe)\|_2 \le \|Pe\|_2.
\]
Finally, since $\mathrm{Var}(e)=\frac1n\|Pe\|_2^2$, we obtain
\[
\mathrm{Var}(e') \le \mathrm{Var}(e),
\]
i.e., the dispersion of neighborhood averages cannot increase under such mixing. This
formalizes the variance-reduction intuition: higher clustering creates overlapping neighborhoods,
which stabilizes local averages and attenuates extreme perceived deviations.
\hfill$\square$}

\subsection{Discussion}

Taken together, these results show that perceived discrimination is shaped by
multiple, distinct topological features.
Homophily and degree heterogeneity amplify perceived unfairness by distorting
local exposure, while clustering mitigates dispersion by stabilizing
neighborhood-level observations.
These mechanisms operate independently of population-level fairness and explain
why objective parity may coexist with persistent perceived discrimination.

%-----------------------------------
\section{Numerical Illustration and Discussion}\label{sec:6}
%-----------------------------------

This section provides a numerical illustration of the mechanisms identified in
Sections~3--5 and discusses their interpretation and implications.
The simulations are intended as qualitative illustrations rather than exact
tests of the theoretical results.
\subsection{Numerical illustration}\label{sec:6.1}

We simulate networks with $n$ nodes partitioned into two groups $A$ and $B$ with proportions
$(\pi_A,\pi_B)$.
Edges are generated according to a two-group stochastic block model (SBM): conditional on group labels,
pairs $(i,j)$ are connected independently with probability $p_{\mathrm{in}}$ if $S_i=S_j$ and
$p_{\mathrm{out}}$ if $S_i\neq S_j$.
We parameterize homophily by $\rho\in[0,1)$ by setting
$p_{\mathrm{in}}=p(1+\rho)$ and $p_{\mathrm{out}}=p(1-\rho)$ (Section~\ref{sec:5}),
where $p>0$ controls the overall density.
We vary $\rho$ on a grid in $[0,\rho_{\max}]$ and generate multiple independent graph realizations for each value. Unless stated otherwise, we fix $n=400$ and $(\pi_A,\pi_B)=(0.5,0.5)$, and we use $R$ independent graph realizations per value of $\rho$.

\smallskip
\noindent\textbf{Assigning acceptance probabilities.}
\responseplus{For each simulated graph, each node $i$ is assigned an acceptance probability $h(i)\in[0,1]$
according to one of the following scenarios:
\begin{itemize}
\item \emph{Group-based rule:} $h(i)$ depends only on $S_i$ (e.g., $h(i)=h_A$ on $V_A$ and $h(i)=h_B$ on $V_B$).
\item \emph{Degree-based rule:} $h(i)$ depends only on the node degree $d_i$ through a monotone mapping
(e.g., a normalized degree score in $[0,1]$).
\item \emph{Mixed rule:} $h(i)$ combines a group component and a degree component. Specifically, we set
\begin{equation}\label{eq:sim_mixed_rule}
h(i)=\Pi\!\left(\alpha\,h^{\mathrm{grp}}(i) + (1-\alpha)\,h^{\mathrm{deg}}(i) + \varepsilon_i\right),
\end{equation}
where $\alpha\in(0,1)$, $\Pi(x)=\min\{1,\max\{0,x\}\}$ clips to $[0,1]$, and $\varepsilon_i$ is a small mean-zero noise term.
In the experiments we take $\alpha=0.7$.
We instantiate the group component by drawing
$h^{\mathrm{grp}}(i)\sim \mathrm{Beta}(4,2)$ if $S_i=A$ and $h^{\mathrm{grp}}(i)\sim \mathrm{Beta}(2,4)$ if $S_i=B$,
and we set $h^{\mathrm{deg}}(i)$ to be an increasing normalized function of $d_i$ (e.g., rescaled ranks or $d_i/\max_j d_j$).
\end{itemize}
Unless stated otherwise, demographic parity is \emph{not} imposed exactly in these simulations (so the global gap may be nonzero).}

\begin{figure}[t]
  \centering
 \includegraphics[width=\linewidth]{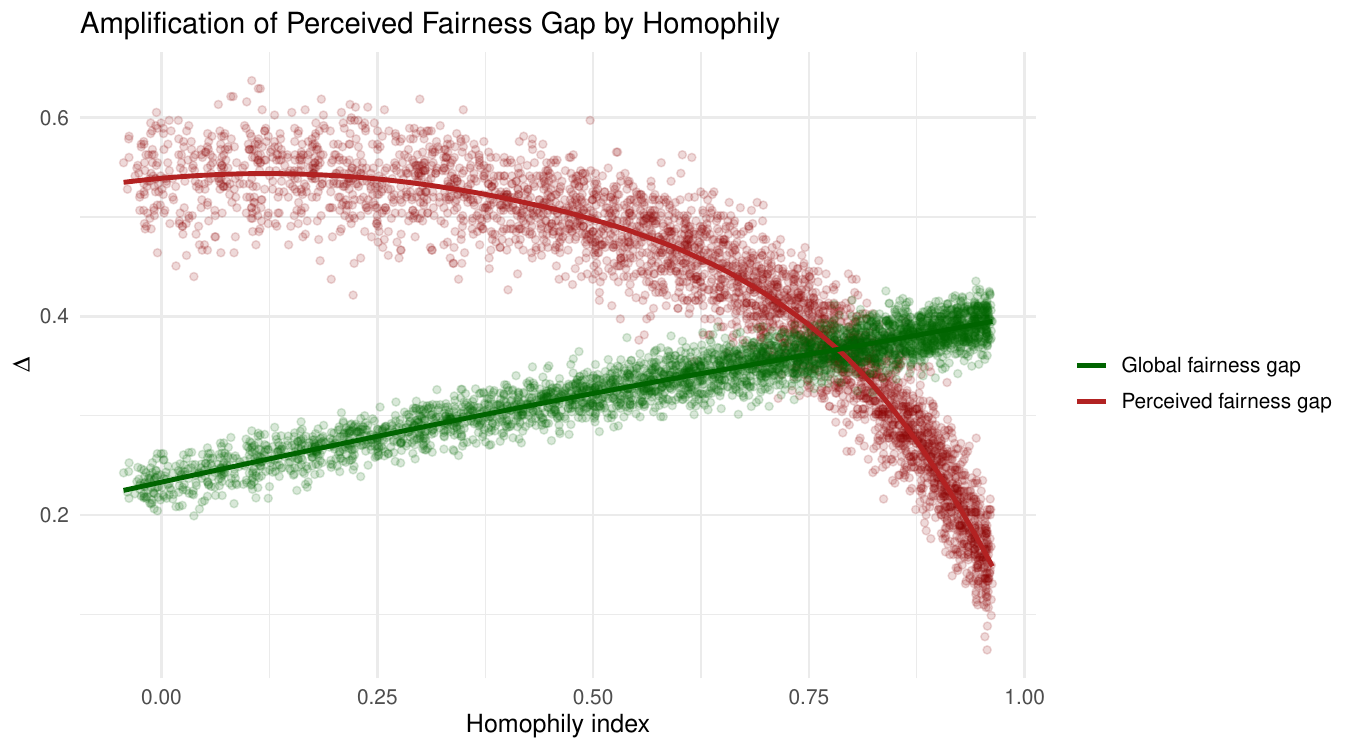}
  \caption{%
  Perceived and global fairness gaps as functions of the homophily index $\rho$.
  Each point corresponds to one network realization; solid lines show LOESS
  smoothing. For small to moderate homophily, perceived unfairness increases
  approximately linearly, while non-monotonic behavior may arise at high
  homophily due to neighborhood homogenization.}
  \label{fig:homophily_simulation}
\end{figure}

\smallskip
\noindent\textbf{Quantities reported.}
\responseplus{For each simulated network we compute:
(i) the \emph{global} (probability-based) fairness gap
\[
\Delta_{\mathrm{global}}(h):=\frac{1}{|V_A|}\sum_{i\in V_A}h(i)-\frac{1}{|V_B|}\sum_{i\in V_B}h(i),
\]
and (ii) the \emph{perceived} fairness gap at visibility radius $r=1$,
$\Delta^{\mathrm{prob}}_1(h)=\mathrm{Vis}^{\mathrm{prob}}_1(A;h)-\mathrm{Vis}^{\mathrm{prob}}_1(B;h)$
(Equation~(6)).
Results are displayed across realizations; in Figure~\ref{fig:homophily_simulation}, each point corresponds to one network realization and the solid line is a LOESS smoother.}

\paragraph{Main observation.}
Figure~\ref{fig:homophily_simulation} plots the perceived and global fairness
gaps as functions of the homophily index $\rho$.
\responseplus{For small to moderate values of $\rho$, perceived unfairness
increases approximately linearly with homophily, consistent with the
first-order expansion in Proposition~\ref{prop:homophily}.}
At larger values of $\rho$, the perceived gap may decrease as neighborhoods
become nearly homogeneous, reducing cross-group comparisons.
This non-monotonic behavior does not contradict the theoretical results, which
characterize local behavior around $\rho=0$.

% \responseminus{The simulations confirm that perceived unfairness always increases with homophily.}
\responseplus{In summary, the simulations are consistent with a positive local dependence on $\rho$ (as captured by Proposition~\ref{prop:homophily}), but also show that at high homophily the perceived gap may saturate or decrease.}

\subsection{Interpretation and policy implications}

We now discuss how the proposed framework should be interpreted and how it may
inform policy analysis.

\paragraph{Perceived versus objective fairness.}
Perceived unfairness is not a substitute for objective fairness criteria such as
demographic parity or equality of outcomes.
Rather, it captures how individuals experience fairness through local social
comparisons.
As shown in Sections~4 and~5, objective parity may coexist with substantial
perceived unfairness when network structure distorts local exposure, and
conversely low perceived unfairness may arise in highly segregated settings
despite persistent demographic disparities.

\paragraph{Why perceptions matter.}
Perceptions of fairness influence behavior, trust, and participation in social
and economic systems.
Even when decisions satisfy formal fairness constraints, persistent perceived
unfairness may undermine legitimacy or compliance.
Our results provide a structural explanation for such mismatches by showing how
network topology shapes local observations.

\paragraph{Policy objectives.}
\responseplus{Policy objectives concerning outcomes and perceptions are
conceptually distinct.}
For realized outcomes, the goal is to reduce disparities in expected values
while improving outcomes for all groups.
For perceptions, the goal is to reduce systematic differences in how fairness is
experienced locally.
Our framework is intended to inform the latter objective by identifying how
network features such as homophily, degree heterogeneity, and clustering affect
perceived fairness.
It does not advocate relaxing outcome-based fairness constraints.

\paragraph{Limits and extensions.}
Finally, we emphasize that reducing perceived unfairness through segregation or
information restriction is not a desirable policy solution.
While such mechanisms may mechanically reduce perceived gaps, they do so by
limiting exposure rather than addressing underlying inequalities.
Future work could extend the present framework to attribute-based homophily,
dynamic networks, or learning processes, and study how perception and behavior
co-evolve over time.

%-----------------------------------
\section{Discussion and Extensions}\label{sec:7}
%-----------------------------------

This paper studies perceived fairness as a network-dependent phenomenon.
While objective fairness criteria are defined at the population level,
individuals evaluate fairness through local comparisons shaped by network
structure.
Our results show that these two perspectives may diverge systematically.

\paragraph{Interpretation.}
Perceived unfairness is not a substitute for objective fairness.
Low perceived unfairness may coexist with substantial demographic disparities
in segregated networks, while perceived unfairness may persist even under
demographic parity.
The framework highlights this potential misalignment without advocating any
normative trade-off.

\paragraph{Policy implications.}
Policy objectives concerning outcomes and perceptions are distinct.
Reducing disparities in expected outcomes remains essential, but addressing
persistent perceived unfairness may require complementary interventions that
modify exposure or information aggregation.
Network structure thus plays a critical role in how fairness is experienced.

\paragraph{Extensions.}
Several extensions are natural.
Future work could study attribute-based homophily, dynamic or adaptive
networks, and learning processes in which perceptions feed back into behavior.
Another direction is to couple perceived fairness with endogenous network
formation.

\paragraph{Conclusion.}
Perceived fairness is a structural property of networked systems.
By formalizing how topology shapes local comparisons, this paper provides a
unified framework to study the gap between objective and perceived fairness in
social and algorithmic networks.

%-----------------------------------
\section{Conclusion}\label{sec:8}
%-----------------------------------
We proposed a mathematical framework linking network structure and perceived fairness.
Our analysis highlights how local perception can deviate from global fairness even when algorithms are unbiased in aggregate.
Future research will connect these theoretical insights with empirical data from collaborative or decentralized systems.

\bibliographystyle{elsarticle-harv}
\bibliography{biblio}

\end{document}